# Certifying the intrinsic character of a constitutive law for semi-crystalline polymers: a probation test.

**S. André[1] . S. Becker[1], C. Noûs[2], L. Farge[1], A.Delconte[3]**

E-mail: _stephane.andre@univ-lorraine.fr_

(1) _LEMTA – CNRS UMR 7563 - Université de Lorraine, 2 avenue de la Forêt de Haye, 54505, Vandoeuvre-Lès-Nancy, France_

(2) _Cogitamus Laboratory_

(3) _CRAN – CNRS UMR 7039 - Université de Lorraine, 2 avenue de la Forêt de Haye, 54518, Vandoeuvre-Lès-Nancy, France_

**Abstract**: A study of methodological nature demonstrates the efficiency of a probation test allowing for the intrinsic character of a rheological constitutive law to be assessed. Such a law is considered here for Semi-Crystalline Polymers exhibiting necking and for large deformation. In the framework of a $(\dot{\sigma}, \sigma, \dot{\varepsilon}, \varepsilon)$ behavior's law, tensile experiments conducted at an imposed constant strain rate $\dot{\varepsilon}_0$ bring true stress responses from which constitutive (material) parameters can be identified from Model-Based Metrology concepts. The same experiment repeated at various strain rates gives then access to the dependence of the non-elastic parameters on the strain rate. Then the intrinsic law is tested severely by considering a new set of experiments carried out for constant displacement rates of the grips. In that case, the specimens show local strain rates which evolve strongly during the test (by a factor of 5-10 here). The parameter identification process requires then the introduction of the exact realized input strain and strain-rate command into the model. Accounting for strain rate dependency requires additionally the knowledge of the preliminary identified strain rate dependence of the non-elastic constitutive parameters for good predictions of the experimental response directly. This is what is proven here. The conclusion speaks in favor of a possible upgrade of international standards for the mechanical characterization of polymers based on constant strain-rate tensile tests and properly applied model-based metrology.

## 1. Introduction

Model-Based Metrology (MBM) means essentially a whole of procedures, at the heart of which are the principles of parameter estimation theory (identifiability proofs) to treat data with the objective of obtaining measurable quantities of physical interest. Such an approach should be a rule of thumb each time a model is used in conjunction with observed quantities because (i) it allows to state in full transparency the performance of the model and the quality of the estimates, but, most important, (ii) it gives a feedback of possible missing foundations of a model or theory. These are revealed by the shape of after-identification residuals (the gap curve between observed and model-reconstructed signals). Such practices are very common in other fields of science (especially in the heat transfer community which was at the origin of the International Conference on Inverse Problems in Engineering ICIPE, but also of course in the automatic and system identification communities) and many textbooks on this subject are now available (Beck and Arnold 1977; Walter and Pronzato 1997; Aster et al. 2013 to cite a few references). But it is pretty much absent from mechanical/rheological studies. In this field, an example of the strength offered by these practices can be studied in Maillet et al. (2013) where it is shown how two models equally optimal for experimental data description can be discriminated through mathematical arguments of parameter identifiability. In the paper of Blaise et al. (2016), these latter were used to characterize the behavior's law of a semi-crystalline polymer (SCP). It was shown in particular that the instantaneous elastic modulus can be retrieved from the common tensile test with a very good agreement with the values determined more directly by other techniques operating at microscopic level. The present paper aims at giving now the second step of the methodology. It consists in proving the validity of a given constitutive law model with a probation test based on its underlying physical concepts: if a rheological relationship $(\dot{\sigma}, \sigma, \dot{\varepsilon}, \varepsilon, \beta)$ is really intrinsic to some material then, any loading path producing strain and strain rate fields heterogeneities in the specimen structure should lead to the same set of model parameters $\beta$. In section 2, information regarding the constitutive model we used and the numerical computation associated within the identification process is given. Section 3 describes the three test cases used for this study. Section 4 discusses aspects of the identifiability of material parameters in the most convenient case where tensile experiments are conducted with the specimen center constrained to follow a constant strain rate path. It additionally produces a so-called 'companion' relation of the constitutive law, in charge of describing strain rate effects on the relaxation spectrum. Section 5 presents the direct computations and identification results that were obtained from a probation test: tensile experiments performed now at imposed displacement rate, where strains and strain rates evolve now in a totally different manner with respect to time.

## 2. Constitutive model and numerical treatment

### 2.1. Operational set of equations

The thermodynamical foundations of the model have been given elsewhere (Cunat 1991 2001a; André et al. 2012) and we focus directly on the mathematical operational form which consists in a set of modal Ordinary Differential Equations (ODE). Each mode is denoted by the $j$ subscript in equation (1) and the overall stress response $\sigma(t)$ is obtained by summation over all modes, meaning that each mode contributes to the global response. Each modal ODE contains the simple description of a relaxation process through a first order kinetic model of characteristic/modal time $\tau_j$. This can be seen in a first approach as a set of analogical Voigt-Maxwell units which would be connected in parallel. These units are uncoupled one with each other (in the simplest view), but the weight $p_j^0$ of a single unit on the global response is connected to the modal time $\tau_j$ through a single "universal" law. As a result and for a given spectrum distribution, a hierarchically recursive scheme applies to the modal weights. More details on this are available in André et al. (2003), where the connection of this approach to non-integer differential constitutive laws is explained (see for example Lion 1997).

$$\dot{\sigma} = \dot{\sigma}^u + \dot{\sigma}^d = \sum_{j=1}^{N} \dot{\sigma}_j = \sum_{j=1}^{N} \left( E_j^u \, \dot{\varepsilon} - \frac{\sigma_j - \sigma_j^r}{\tau_j} \right) \tag{1}$$

In equation (1), $E_j^u$ corresponds to the $j^{th}$ modal unrelaxed (or instantaneous) modulus. $E_j^u$ and $\sigma_j^r$ can be defined by $E_j^u = p_j^0 E^u$ and $\sigma_j^r = p_j^0 \sigma^r$ with the weighting coefficients $p_j^0$. $E^u$ stands for the common elastic modulus (Young modulus denomination will be avoided to favor the thermodynamic foundation of this parameter as the twice differentiated thermodynamic potential with respect to the strain variable). Therefore we have the modal weights fulfilling the normalized condition:

$$\sum_{j=1}^{N} p_j^0 = 1 \tag{2}$$

$\sigma^r$ refers to the stress in the relaxed state, a thermodynamic state which has been properly defined by Prigogine and Defay (1958). It corresponds to the stationary state for non-equilibrium internal forces and is defined by $\dot{A}^r = 0$ where $A$ denotes the affinity in the T.I.P. theory (De Donder 1936). It implies formally a direct coupling between internal variables reflecting the microstructure evolution and the macroscopic command (strain rate here). Logically, this state can be identified when all viscoelastic modes get through, which we associate to the set-up and further evolution of the fibrillar state in SCP. It can then be simply and very efficiently modeled here with the hardening law

$$\sigma^r(t) = G \, \varepsilon_{HT}(t) \tag{3}$$

suggested by theoretical considerations on the elastomeric state (Treolar 1975; Haward 1993 2007; Arruda and Boyce;, 1993; Tervoort and Govaert 2000) and experimental evidences (Supplementary Information Fig. S1). The parameter $G$ (called the hardening modulus) introduces the proportionality of the relaxed stress with the 'Haward-Thackray' strain variable $\varepsilon_{HT} = \lambda^2 - \lambda^{-1} = \exp(2\varepsilon) - \exp(-\varepsilon)$ where $\lambda$: extension ratio, and $\varepsilon$: logarithmic true strain.

Finally, a spectrum of $N$ relaxation times is chosen as logarithmically distributed over a determined number of decades $d$ below a maximum relaxation time $\tau_{max}$.

$$\tau_j = \tau_{max} \, 10^{-\left(\frac{N-j}{N-1}\right)d} \tag{4}$$

Generally, $N = 50$ dissipative modes are considered to figure out a continuous spectrum. A number of decades $d = 6$ is generally required for the tested polymers so that the spectrum extends to sufficiently low relaxation times beyond which the model becomes unsensitive. The spectrum is then determined once the single parameter $\tau_{max}$, the longest relaxation time, is fixed.

The key conceptual tool is to connect the dissipative modal weight $p_j^0$ to its corresponding relaxation time $\tau_j$ and this is made with a thermodynamically based argument stating an equipartition of the entropy created by each dissipative (relaxation) mode (Cunat 1991 2001; Faccio-Toussaint et al. 2001). This leads to

$$p_j^0 = \frac{\sqrt{\tau_j}}{\sum_{j=1}^{N} \sqrt{\tau_j}} \tag{5}$$

The constitutive relationship expressed through equation (1) can be rewritten now as

$$\dot{\sigma} = \sum_{j=1}^{N} \dot{\sigma}_j = \sum_{j=1}^{N} \left( p_j^0 E^u \dot{\varepsilon} - \frac{\sigma_j - p_j^0 G \varepsilon_{HT}}{\tau_j} \right) \qquad (6)$$

The functional form associated to this behavior's model can be written as $(\dot{\sigma}, \sigma, \dot{\varepsilon}, \varepsilon, \beta)$ and requires the knowledge of the following parameter vector $\beta = \left[ E^u, \tau_{max}, G \right]$. The relevance of this 3-parameter model has been proven in Blaise et al. (2016). The sensitivity analysis and parameter identification procedures have shown how well conditioned is this model with respect to the Parameter Estimation Problem (PEP). The analysis of the after-identification residuals proved that the viscoelastic behavior is correctly caught by the modal recursive approach, as well as the hardening stage within the approach of Eq.(3). Evidence of this will appear again when presenting identification results in Section 4.

2.2. Numerical solution

When considering a tensile test at constant imposed strain rate $\dot{\varepsilon}_0$ (i.e. $\varepsilon(t) = \dot{\varepsilon}_0$), this system of very simple ODE's with simple forcing term (input excitation) can be solved analytically, either directly or using the Laplace transform. The analytical solution is:

$$\sigma_j(t) = \dot{\varepsilon}_0 \left[ p_j^0 \tau_j E^u \left( 1 - e^{-t/\tau_j} \right) + G \left( \frac{3p_j^0 \dot{\varepsilon}_0 e^{-t/\tau_j} - p_j^0 \left( 2\dot{\varepsilon}_0 + 1/\tau_j \right) e^{-\dot{\varepsilon}_0 t} + p_j^0 \left( 1/\tau_j - \dot{\varepsilon}_0 \right) e^{2\dot{\varepsilon}_0 t}}{\tau_j \dot{\varepsilon}_0 \left( 1/\tau_j - \dot{\varepsilon}_0 \right) \left( 2\dot{\varepsilon}_0 + 1/\tau_j \right)} \right) \right]$$

for each modal "branch" and $\qquad (7)$

$$\sigma(t) = \sum_{j=1}^{N} \sigma_j(t) \text{ for the overall response}$$

Tensile tests at constant imposed strain rate are, for this major reason but among other ones, a clear advantage when metrology is the foremost objective.

In the case considered later where the ever changing true strain rate excitation must be introduced as the input of the model, a numerical scheme is required that computes the solution as time steps proceed. We use then a classical two steps implicit Euler Backward second order scheme (GEAR or bE2) applied on the system of ODE's of (Eq.6). For the $j^{th}$ modal component, we consider

$$\dot{\sigma}_j(t) = f\left(t, \sigma_j\right) = -\frac{\sigma_j(t)}{\tau_j} + p_j^0 E^u \dot{\varepsilon}(t) + \frac{\sigma_j^r(t)}{\tau_j} \qquad (8)$$

discretized through

$$\sigma_j^{n+1} = \frac{4}{3}\sigma_j^n - \frac{1}{3}\sigma_j^{n-1} + \frac{2h}{3} f\left(t^{n+1}, \sigma_j^{n+1}\right) \qquad (9)$$

with the time step $h = t^{n+1} - t^n$.

Initialization of the algorithm rests upon the two first time steps: the initial condition $\sigma_j^0$ (material in equilibrium state) and a first single step Euler Backward scheme of order 1. An alternative scheme has been used, very precise too, based on a semi-analytical approach of the integration of the system of eqs. (6). This algorithm (SA3) is described in Sorvari and Hämäläinen (2010). It is used in Section 4 to ensure proper implementations of both algorithms but it extends pretty much the computation times. Furthermore, in the case of a non-idealized forcing term (experimental data), it amplifies the input noise and appears therefore not appropriate to this study.

## 3. Considered test cases

Three experiments will be considered for this analysis that have been obtained for specimens of HDPE (same original material from Röchling Grade Natural) and all at same room temperature. Reproducibility is not discussed as already proved excellent with our metrological equipments (elements provided in Blaise et al., 2010, Ye et al., 2015). Important elements about the metrological tools involved in these experiments are not discussed here but also available from the above-mentioned references.

**Experiment 1** refers to a tensile test performed at an imposed constant strain rate $\dot{\varepsilon} = 0.005\ s^{-1}$. It relies on a VideoExtensometer working with a limited number of markers located in the specimen gauge, that provides a real-time measurement of the local strains. This signal drives the actuator of a hydraulic tensile machine through a PID feedback loop. In the case of a ramp input; the realized strain signal perfectly follows the input command as can be seen in Fig.1: the constant strain-rate is perfectly realized with a small random noise of 0-mean superimposed to it.

**Experiment 2** refers to a tensile test performed at an imposed constant displacement rate $\dot{u} = 0.02 \, mm/s$ on the same machine.

**Experiment 3** refers to a tensile test performed in same conditions as experiment 2 but with a 4 times greater imposed constant displacement rate $\dot{u} = 0.08 \, mm/s$.

For all 3 experiments, a 3D (stereo) DIC system provides strain measurements on the 2 front and lateral faces of the specimen gauge, allowing for precise true stress measurements (Farge et al., 2015). These 3 experiments will serve the objective of the demonstration. With respect to an intrinsic behavior's law expressed in terms of true stress and strain, they provide very different strain and strain rate excitations on the Representative Element Volume (REV) of the material as shown in Figure 1.

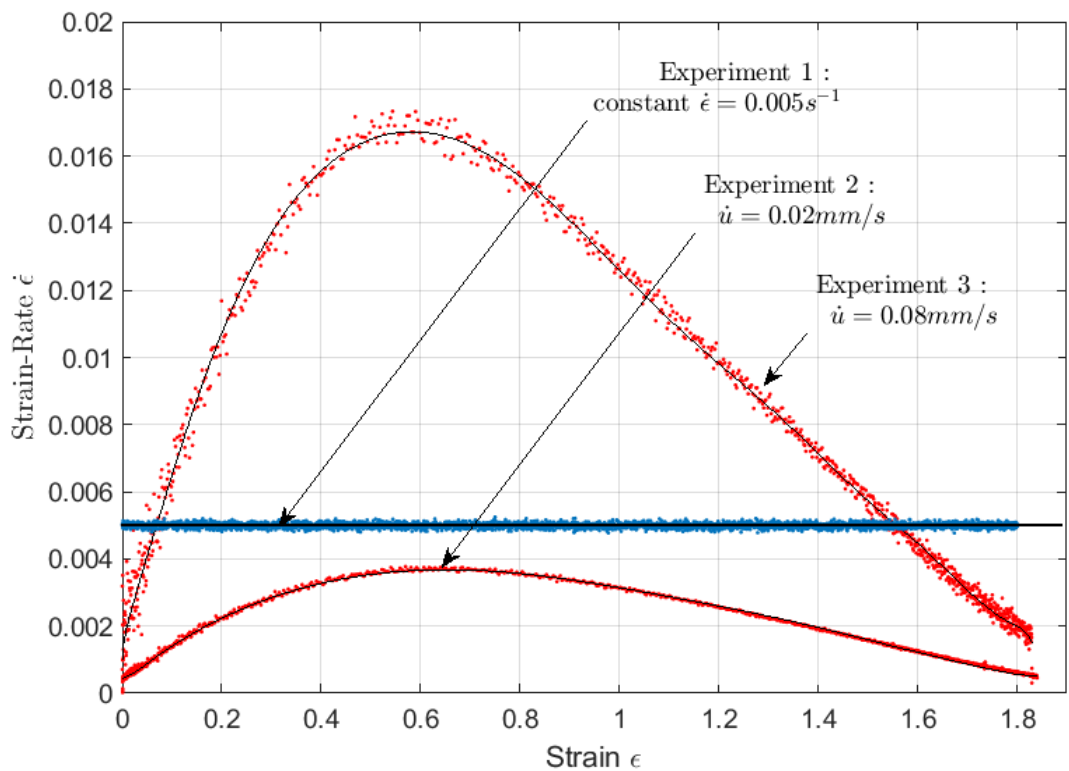

Figure 1: Strain-rate evolution with strain during a tensile test performed at constant imposed strain or displacement rates.

## 4. Experiment 1 (constant strain rate): Identification of model parameters.

Applying the model (Eq. 6) to the experimental data obtained for experiment 1 through an inverse estimation of the model parameters will give us key elements of discussion to assess our demonstration. Estimation is performed on a least-square criterion optimization and using indifferently a Levenberg-Marquardt or simplex algorithm. Because the problem is relatively well-conditioned (see explanations in Blaise, 2016) results are always identical. Accounting for the Levenberg-Marquardt algorithm's fast computation times, this latter will be used preferentially.

In figure 2 below, we show a typical result of the adjustment of the model to the data of experiment 1. We selected a special set of data which, unlike the one shown in Fig.1, suffers from accidental superimposed high-frequency noise, an artefact due to some un-desired vibrations. This signature will make clearer our analysis of the residuals obtained after the optimization process. Four different identifications were applied on these data whether the true or idealized strain and strain-rate were considered in the model or the numerical or analytical versions of the model were used (Eq. (7) versus eq. (9)).

- Identification 1 = analytical model of Eq.7, which implies fully idealized ramp command effectively realized in the test.
- Identification 2 = Numerical algorithm (bE2) which uses the real (noisy) strain and strain-rate input signals. Considered as the reference case for the study.
- Identification 3 = Numerical algorithm (bE2) which uses "filtered" (un-noisy) strain and strain-rate input signals. Typically, a linear regression is applied to the measured input strain command and is substituted for it in the algorithm.
- Identification 4 = Same as Identification 3 but using the (SA3) numerical algorithm to prove consistency.

As shown by Fig.1, the least-squares optimization of the model produces absolutely identical results although the identified parameter values can show 2-3% of variations (see Table 1). The model can explain the data very well: the after-identification residuals -i.e. discrepancy between experimental and reconstructed data- produce a control of the quality of the optimization process. In the residuals signature, two components are always present that can be recognized easily in Fig.2.

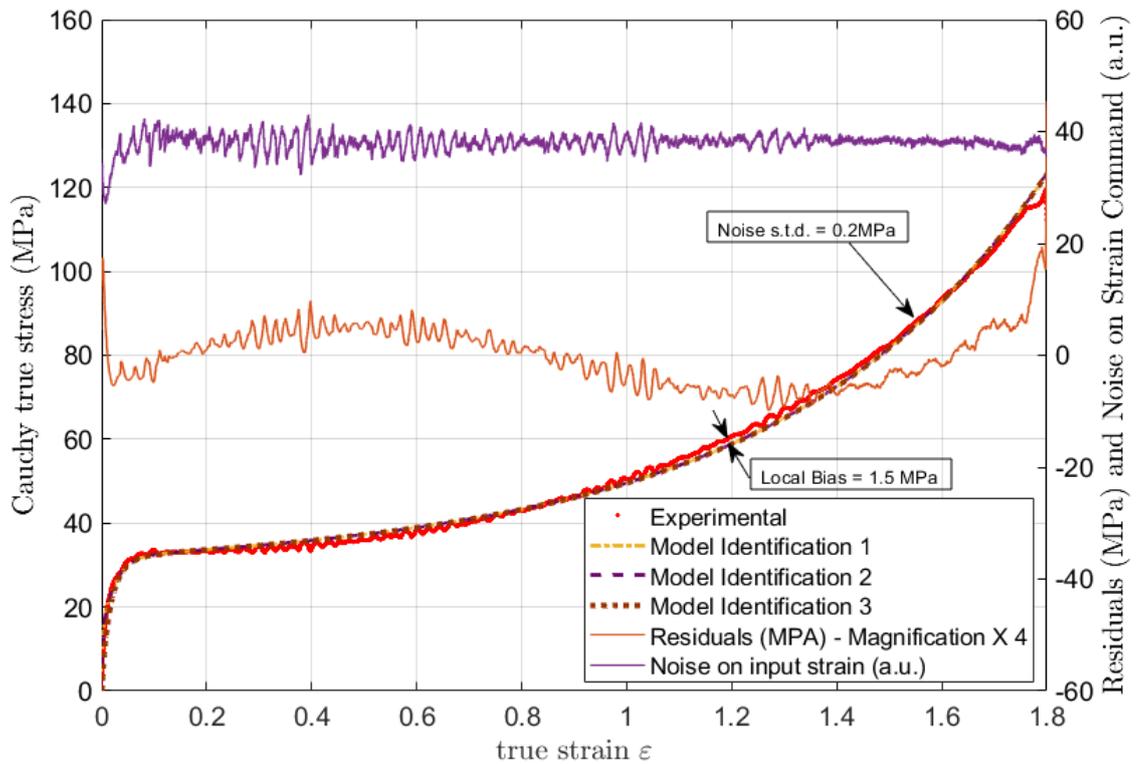

Figure 2: True stress-true strain curves for **Experiment 1** – Left axis: Experimental data and adjusted model; Right axis: After-identification residuals (magnified by a factor of 4) and difference between input and realized strain ramp command (arbitrarily amplified and shifted to a mean value set at +40).

- The noise at higher frequency is a direct contribution of the initial noise pre-existing on the input excitation (in general it is essentially the noise of the sensors and possible electronic conditioning of their raw signal). Estimation parameter theory proves that it affects the variance on the estimated parameters but if the model is a perfect idealization of the experiment and if the sensitivity analysis shows highly sensitive and perfectly un-correlated parameters, it does not bias the estimated parameters and is fully recoverable in the identification residuals (i.e. the s.t.d of the residuals should be equal to the s.t.d. of the noise on the data). This is nearly the case here. After-identification residuals (right axis) are nearly centered on a 0-mean and corrupted by a "high" frequency signal which corresponds exactly to the 0-mean input noise observed on the strain signal. This can be seen from the data set shifted to the arbitrary 40 mean value for clarity of Fig. 2 (second curve, same right axis) which indeed corresponds to the direct difference between the input measured strain and its idealized ramp behavior.

- The long range wavy behavior of the residuals is reflecting the bias due to some deficiency in the model/experiment agreement. This bias either results from an uncontrolled bias in sensor measurements but in contemporary science is more likely the result of a more or less pronounced incompatibility between the assumptions of the model (an ideal design) and the objective experimental conditions. From our insight on this specific problem, it is more likely due to the approximate phenomenological description of the relaxed state according to eq.(3). A bias due to the force signal conversion into a true stress signal, based on DIC surface measurement strains on the three surfaces of the specimen in its central part cannot be excluded totally either. We indicate on Fig. 2 where the local bias is at maximum (of the order of 1.5MPa at a strain of 1.2).

| | $\hat{E}_u$ (MPa) | | $\tau_{max}^T$ (s) | | $\hat{G}$ (MPa) | |
|---|---|---|---|---|---|---|
| | | Relative error | | Relative error | | Relative error |
| Identification_1 *Analytical model (Eq.7)* | 3112 | -3.5% | 5.394 | +3.6% | 2.552 | -0.2% |
| **Identification_2** ***Reference Numerical algorithm bE2 (Eqs 8 and 9)*** | **3224** | **-** | **5.207** | **-** | **2.556** | **-** |
| Identification 3 *Numerical algorithm bE2* | 3215 | -0.3% | 5.22 | -0.25% | 2.55 | 0% |
| Identification_4 *Numerical algorithm SA3* | 3218 | -0.2% | 5.22 | -0.25% | 2.557 | 0% |

Table 1: Identified model parameters for the different options of model/algorithm for the identification.

Of course it is essential to constraint the bias as low as possible. In general the variance of the noise on the true strain signal is far below the bias, and is of the order of $0.04 \, Mpa$ (in our experiments, which corresponds to a noise on the Force sensor of s.t.d less than 2N).

In Table 1, the identification 2 using bE2 algorithm has been taken as a reference as a numerical approach is mandatory to treat data obtained in the case of Experiments of type 2 and 3.

## 5. Experiments at constant displacement rate: the probation test.

Because a tensile test performed at <u>constant displacement rate</u> makes the central part of the specimen (where the true stress-true strain curve is recorded) subjected to <u>very different strain rates</u> (Fig.1) due to necking, the constitutive model which is expressed in terms of a $(\dot\sigma, \sigma, \dot\varepsilon, \varepsilon)$ law must obviously take into account the real excitation $\dot\varepsilon$ but not only. The key point is that under evolving strain rate excitations, the specimen's viscoelastic response reflects a change in relaxation times. As a result, the condition for the model to be applied to fit any kind of experimental excitations with pertinent results, is to have available a relationship between the strain rate and -in the framework of our model- the maximum identified relaxation time. This latter was obtained in Blaise et al. (Blaise et al., 2016), the constitutive law having been applied to a series of stress-strain curves monitored at different constant strain rates. Thanks to the well designed and parcimonious model described in section 2 through Eq.(7), the identified parameters show a strain rate dependence which is expected to be a real intrinsic property of the material. For the seek of some universal character, the Weissenberg number $We$ was introduced. It corresponds to the ratio between the time constant $t_{mat}$ characterizing the intrinsic 'fluidity' of the material and the time scale of the experiment or of the observer $t_{exp}$. The fluidity of the material is inversely proportional to the Weissenberg number. In the present case, its maximal value can be calculated by the following formula

$$We = \frac{t_{mat}}{t_{exp}} = \tau_{\max} \; \dot\varepsilon \qquad (10)$$

where $\tau_{\max}$, the maximum relaxation time of the spectrum, is used for the material characteristic time and $1/\dot\varepsilon$ for the experimental characteristic excitation time.

Table 2 summarizes the findings for HDPE.

| Strain rate ($s^{-1}$) | $\tau_{max}$ (s) | $We$ | G (MPa) |
|---|---|---|---|
| $5.10^{-5}$ | 419.72 | 0.0209 | 1.53 |
| $4.10^{-4}$ | 63.23 | 0.0253 | 1.85 |
| $2,5.10^{-3}$ | 11.02 | 0.0276 | 2.31 |
| $5.10^{-3}$ | 6.01 | 0.0301 | 2.31 |
| $10^{-2}$ | 3.21 | 0.0321 | 2.46 |

Table 2. Average estimated maximal relaxation times $\tau_{max}$, corresponding Weissenberg numbers $We$ and hardening moduli $G$ for the HDPE specimen subjected to tensile tests performed at different strain rates.

### 5.1. Strain-rate dependency of model parameters

- Time spectrum: There are two options regarding the "modeling" of the relaxation times spectrum dependency with respect to strain rate.

  **Option 1** As reported in Table 1, experiments conducted at various imposed strain rates have provided the identification of different maximum relaxation times. The corresponding calculated Weissenberg number varies from more than 50% over 3 decades of strain rates but in an unknown manner (no model available). A plot of $We$ as function of $\dot{\varepsilon}$ in logarithmical scale shows a roughly linear dependency. Applying a regression, gives the following relationship with R-square coefficient of 0.976.

$$We = \tau_{max}^{We} \dot{\varepsilon} = 0.00464 \log(\dot{\varepsilon}) + 0.0367 \qquad (11)$$

When definitely established with prior experiments and if really intrinsic to the material, this relation can be seen as a companion relation to the constitutive law. The spectrum parameter $\tau_{max}^{We}$ can be considered as known (simple division of $We$ by $\dot{\varepsilon}$) once the local strain rate $\dot{\varepsilon}$ imposed locally on the REV is known.

Two controls can be applied to check whether this relationship is valid for our material. Either the adjustment procedure is reduced to the identification of the only one remaining parameter (Relaxed modulus $G$) because the model parameter $\tau_{max}$ is supposed to be known from (11) ($\tau_{max}^{We}(\dot{\varepsilon})$);

Or parameter $\tau_{max}$ will be considered as corresponding to $\tau_{max} = \alpha \tau_{max}^{We}$ and the adjustment procedure will consider both the relaxed modulus $G$ and the correction coefficient $\alpha$ as parameters to identify to measure in some ways, whether or not the residuals can be further minimized when authorizing $\tau_{max}$ to slightly fluctuate around the supposed known value $\tau_{max}^{We}$. Of course, if the relationship (11) perfectly holds, this coefficient should be identified equal to 1. As this relation results from measurements already issued from an identification process, some errors may exist but if rather insensitive to the obtained results, this will prove the intrinsic character of both the constitutive law and its companion material spectrum relation.

**Option 2** A direct fit can be applied to the $\tau_{max}(\dot{\varepsilon})$ data of Table 2. In that case we obtain a perfect representation of the data (R-square coefficient = 1) with the following exponential relation:

$$\tau_{max}^{fit}(\dot{\varepsilon}) = 27.24 \, exp(-2.1 \, log 10^3 \dot{\varepsilon}) \tag{12}$$

The $\tau_{max}^{We}(\dot{\varepsilon})$, $\tau_{max}^{fit}(\dot{\varepsilon})$ and $\tau_{max}(\dot{\varepsilon})$ data are plotted in Figure S1 (Supplementary Information). It is shown that both capture the apparent strain-rate dependence even if the $\tau_{max}^{We}(\dot{\varepsilon})$ approximation underestimates the time spectrum higher bound at very low strain-rates.

- Hardening modulus: In the same way, the Hardening modulus identified from the stress-strain response to an input ramp at constant strain rate evidently depends on the selected strain rate. This is generally observed indirectly by varying the temperature of polymers. The decrease of $G$ when the temperature increases was already observed for both amorphous (Van Melik et al., 2003) and semi-crystalline polymers (Na et al., 2007). It is due to thermally activated relaxation mechanisms that can be evidenced by DMA (Dynamic Mechanical Analysis) measurements on pre-deformed specimens. In the case of polyethylene, the identified relaxation process is the $\alpha$ relaxation mode associated with the slip of mosaic blocks inside the lamellae (Na et al., 2007). Because of the temperature-time equivalence principle, a strain rate increase has the same qualitative effect as a temperature decreases and results in a $G$ increase which is exactly what we observe from parameter $G$ identification from experimental curves. But as far as the authors know, there is no reported idea about the mathematical description of this dependency. Unlike relaxation times which can be assumed to shift rather simultaneously with a change in microstructure, a direct instantaneous dependence on strain rate could hardly be defend for the relaxed modulus $G$. It would be firstly in contradiction with the hypothesis of describing precisely a relaxed state. Secondly, as it describes the elasticity of the SCP's fibrillar network established differently according to the loading path history, it can change physically only in a delayed manner (like for temperature). For those two reasons, we will keep the strict relation (Eq. 3) for the relaxed state, the modulus $G$ representing the apparent hardening elasticity in a somewhat homogenized manner for constant displacement-rate tensile tests.

### 5.2. Overcoming the probation test

The objective of the paper is achieved in this section. Figure 3 shows the best model adjustment and vector $\beta = \left[ E^u, \tau_{\max}, G \right]$ identification on the experimental stress-strain curves obtained for constant displacement rate tensile tests (Experiments of type 2 and 3). The model accounts for the exact realized input strain command with varying strain-rates of Fig.1 (see explanations of the fitting process of these command signals in Supplementary Information: Curve fitting process for the command/excitation signals). But it does not account yet for the strain-rate dependency evidenced in the previous section on the $\tau_{\max}$ model parameter (Eqs.11 or 12). The demonstration is clear. The constitutive law model is unable to reproduce the real behavior which nevertheless has an apparent behavior identical to strain-rate experiments (compare with the stress-strain curve obtained at constant strain-rate in Fig2). The model produces a short time behavior corresponding to a sudden jump in stress with very low yield stress (around 2MPa) and the post-yield behavior is simply unacceptable as well as the identified instantaneous modulus found to be around 60 GPa. An expeditious conclusion would lead to consider the proposed constitutive law as non intrinsic to the material, which is, of course, what a scientist in rheology tries himself to prevent from.

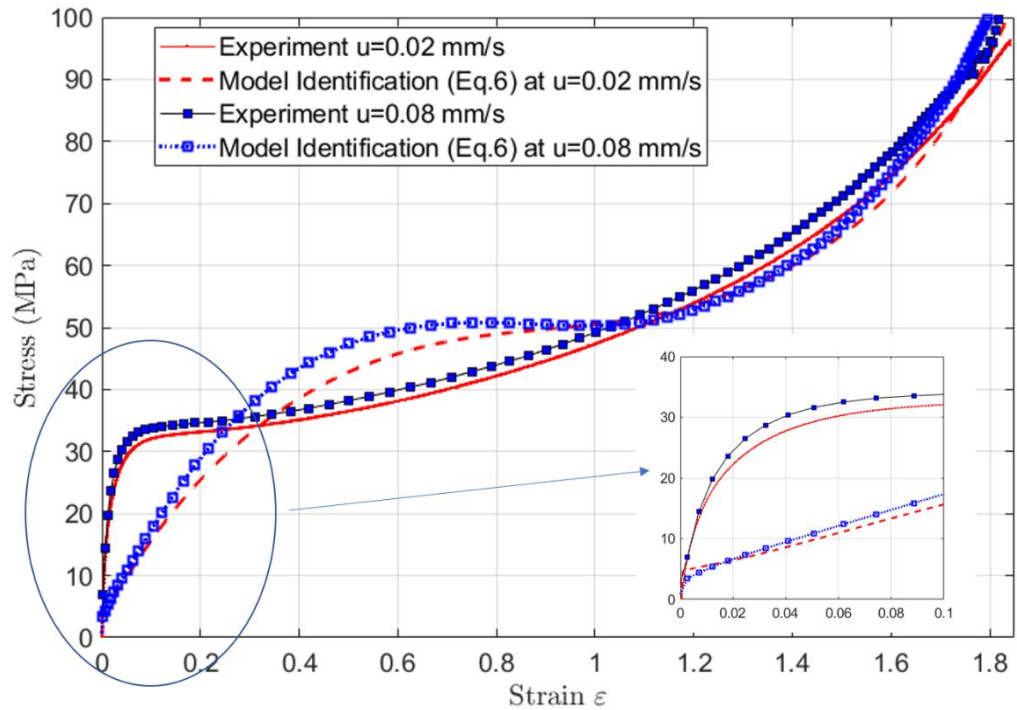

Figure 3: Experiments at constant displacement rate and model inconsistency when intrinsic strain-rate dependency is neglected

If now, the "companion" relation (11 or 12) is used to determine parameter $\tau_{max}$ of the constitutive behavior's law (Eq. 6), then the prediction of the true stress-true strain curves for both experiments at constant displacement-rate (Fig.4a when using Eq.11 and Fig4b when using Eq.12) is very reliable with all rheological regimes of the tensile curve clearly recovered (Comparison to be made with curves of Fig.3). These direct simulations were performed for an instantaneous modulus of $3220 MPa$ (Table 1-Line 2). Concerning the hardening modulus G, the best strategy is to determine it in accordance with the theory used for the relaxed state description as was presented in Blaise et al. (2016). The plot of the experimental stress versus the Haward-Thackray strain variable should exhibit a linear behavior. This is nearly the case for both Experiment2 and Experiment3 tensile tests. From a linear regression (see Figs S2a and S2b in Supplementary Information) it can be easily retrieved as G=1.56 MPa and G=1.64 MPa respectively for the experiments at $\dot{u} = 0.02\ mm/s$ and $\dot{u} = 0.08\ mm/s$ .

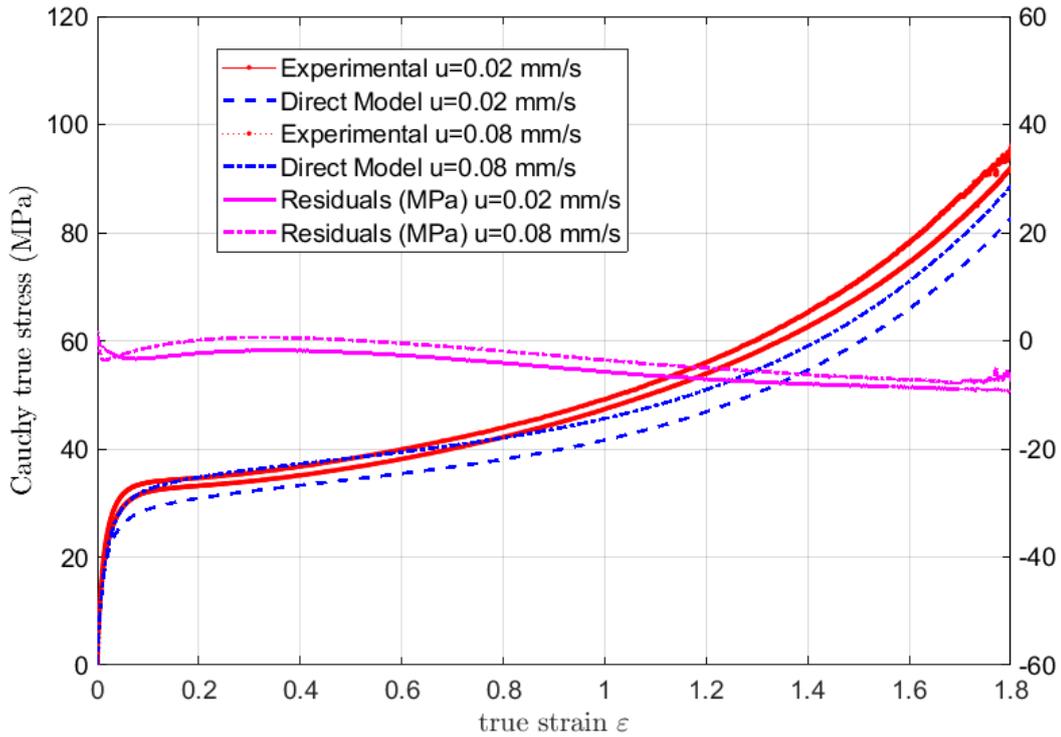

Figure 4a: Comparison between tensile tests at constant displacement rate and direct model computation from eq.6 and companion strain-rate dependency for the relaxation spectrum (Eq.11).

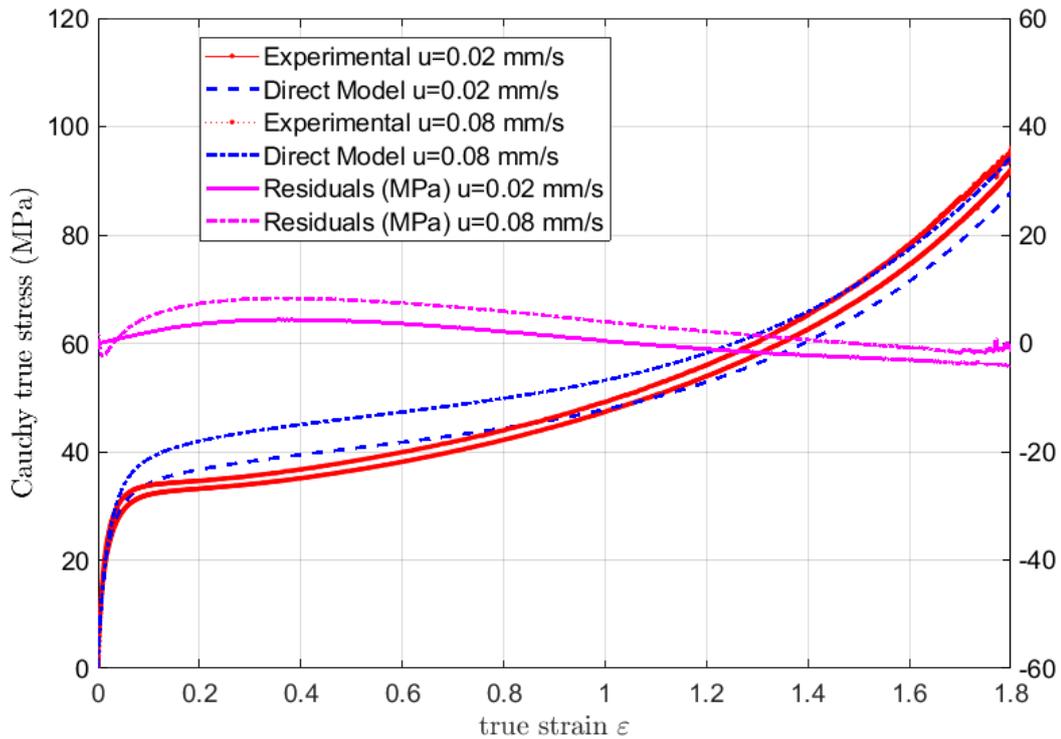

Figure 4b: Comparison between tensile tests at constant displacement rate and direct model computation from eq.6 and companion strain-rate dependency for the relaxation spectrum (Eq.12).

It is clear from this direct modelling that the constitutive model (Eq. 6) is able to describe the experiment in an intrinsic manner, the bias being now enhanced from the consequence of the approximation of Eqs.11 or 12 to determine the spectrum shift with strain-rate variations and the a priori determination of the overall hyper-elastic modulus $G$. The extreme sensibility of the spectrum shift is evident from Figs 4a and 4b. Although the discrepancies between the model prediction and the experiment remain at maximum equal to 10 MPa, it happens rather in the yielding regime (Fig.4b) or in the hardening regime (Fig.4a) whether Eq.12 or Eq.11 are used respectively.

### 5.3. Identification of the model with strain-rate correction

The analysis can be pushed a little forward by producing model adjustments on the experiment, with the same constitutive model of eq.6, the exact realized input command signal but allowing for the adjustment of parameters $\tau_{max}$ and $G$.

Results of identifications are reported in Table 3 for experiment 2 ($\dot{u} = 0.02\ mm/s$). Again and accounting for the results of Table 1, the instantaneous modulus measured in tensile tests of the type of experiment 1 is considered as known and equal to $E^u = 3.22\ GPa$ ('Identifications 2 and 3' in Table 3). In 'Identification2', the parameter $\tau_{max}$ of the relaxation spectrum will be determined as either $\tau_{max} = \alpha \tau_{max}^{We}(\dot{\varepsilon})$ with $\alpha$ allowing for a correction factor to be identified for each test but with the $\tau_{max}^{We}(\dot{\varepsilon})$ assumed relation (Eq.11) or, in 'Identification3', $\tau_{max} = \alpha \tau_{max}^{fit}(\dot{\varepsilon})$ with the $\tau_{max}^{fit}(\dot{\varepsilon})$ assumed relation (Eq.12). Because $\alpha^{ref} = 1$, this will present the advantage of showing directly in % the amount of adjustment made from eqs.11 or 12 to $\tau_{max}$ for providing a better fit. Parameter $G$ will be considered as constant to provide the best overall estimation of the behavior in hardening regime (also it has been observed that the establishment of the fibrillar regime also depends on the strain-rate as discussed earlier). It can be initialized to the estimated value given in section 4.2. In 'Identification4', both three parameters $E^u, \tau_{max}, G$ are identified to check the stability of the estimation process.

| | $\hat{E}_u$ (MPa) | | $\alpha$ (s) | | $\hat{G}$ (MPa) | |
|---|---|---|---|---|---|---|
| | | Relative variation | | Relative variation | | Relative variation |
| **Reference** | **3220** known | - | **1** | | **1.56** (estimated) | |
| **Identification_2** ($\tau_{max} = \alpha \tau_{max}^{We}(\dot{\varepsilon})$) Cost function=1155 | **3220** known | - | 1.11 | +11% | 1.761 | +16% |
| Identification 3 ($\tau_{max} = \alpha \tau_{max}^{fit}(\dot{\varepsilon})$) Cost function=1191 | **3220** **known** | - | 0.928 | -7% | 1.764 | +16% |
| Identification_4 ($\tau_{max} = \alpha \tau_{max}^{We}(\dot{\varepsilon})$) Cost function=1138 | 3428 | +6.5% | 1.045 | +4.5 | 1.765 | +16% |

Table 3: Identifications performed on Experiment 2 (imposed constant displacement rate of $\dot{u} = 0.02\ mm/s$)

Figure 5 shows how well the model identification compares with the experiment for the 3 identification options. The estimated response of the model is the same, the Least-Square cost function leading to the same value at convergence. Because the noise on the experimental strain-stress signals is very low, the residuals (magnified by a factor of 5) are the pure reflect of the remaining bias. It is at maximum equal to 2.5 MPa which is pretty good (compare with Figs 4a and 4b).

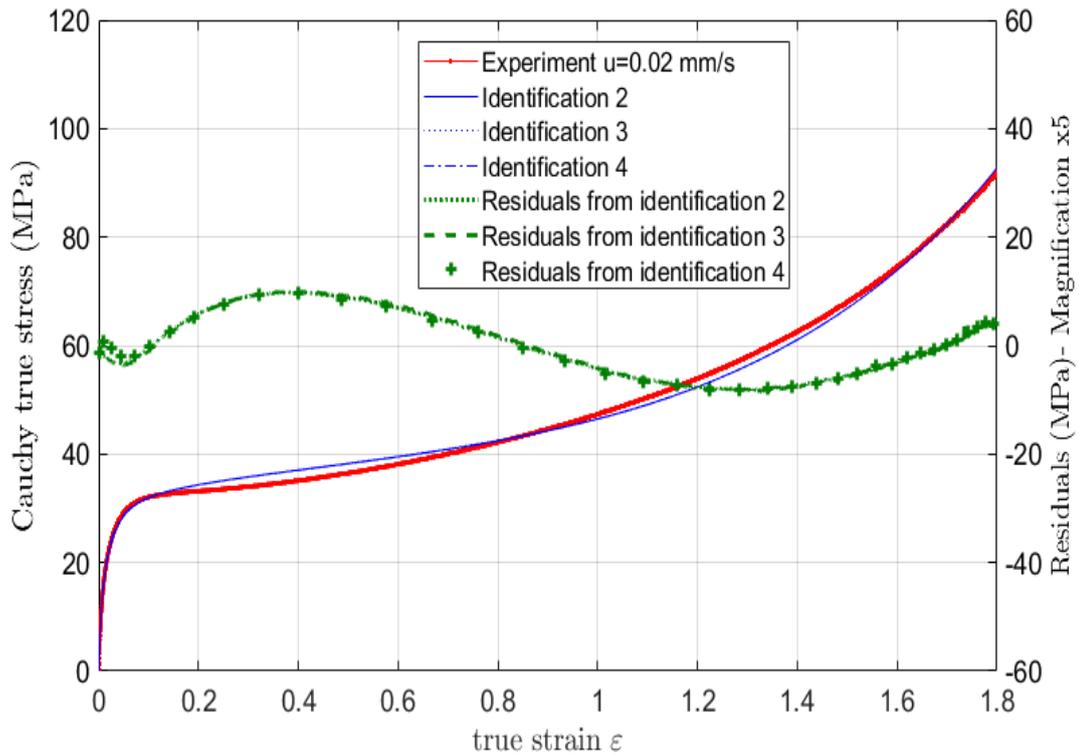

Figure 5: Comparison between tensile tests at constant displacement rate and direct model computation from eq.6 and companion strain-rate dependency for the relaxation spectrum (Eq.12).

Regarding identified parameters, it is shown (Identifications 2 and 3) that whether Eqs 11 or 12 are used for initial estimation of $\tau_{max}(\dot{\varepsilon})$ is not so important. The $\alpha$ value identified to correct (shift uniformly) the relaxation spectrum is of the order of 10% at maximum. When the instantaneous modulus is free to adjust the data, it increases by 6.5% when the $\tau_{max}$ parameter can be less corrected ($\alpha = +4.5\%$). This is just the result of a slight correlation effect existing between these two parameters. Note that assuming no bias in the couple (experiment/model), the ideal relative error on these two parameters was calculated of the order of 2% in Blaise et al. (Blaise, 2016). 'Identification 4' then proves the robustness of the P.E.P. conditioning. Regarding the 'apparent' hardening modulus G, nothing can be concluded except to observe that the initial value estimated from the linear regression at strong deformation is corrected by the estimation process performed on the whole strain range to product better adjustment to the experimental data.

Turning now to the tensile test performed $\dot{u} = 0.08\ mm/s$ (Experiment 3), the same analysis can be made and results of identifications are reported in Table 4. Results concerning the model adjustment, the residuals magnitude (lower than 3.4 MPa), and the parameter estimates are similar to Experiment2 (see Table 4 and Figure 6) except for identification 4. In this case, the instantaneous modulus is not considered as known but as a parameter to estimate. The less good-conditionning of the P.E.P. becomes obvious. Remember that P.E.P. are Non Linear Problems in terms of the parameters (and possibly also the inputs). It is clear here that the strain-rate range in input introduces a stronger correlation between parameters $E^u$ and $\alpha$ (i.e. $\tau_{max}$). These two parameters evolve now strongly (+36% and -24%) and the model adjusts much further than for Identifications 2 and 3: the cost function is reduced by a factor of 25%. This can be seen on Fig. 6 where in that case, the residuals are very minimized at short times, in the early beginning of the tensile test (cross-dot residuals curve of Identification4). The best strategy here is clearly to consider that parameter $E^u$ remains at its known value, previously estimated and to consider only Identifications2 and 3 as reasonable.

| | $\hat{E}_u$ (MPa) | | $\alpha$ | | $\hat{G}$ (MPa) | |
|---|---|---|---|---|---|---|
| | | Relative variation | | Relative variation | | Relative variation |
| **Reference** | **3220** known | - | **1** | | **1.64** (estimated) | |
| Identification_2 ($\tau_{max} = \alpha\tau_{max}^{We}(\dot{\varepsilon})$) Cost function =2100 | **3220** known | - | 1.045 | +4.5% | 1.85 | +12.6% |
| Identification 3 ($\tau_{max} = \alpha\tau_{max}^{fit}(\dot{\varepsilon})$) Cost function=2250 | **3220** known | - | 0.855 | -14% | 1.86 | +12.6% |
| Identification_4 ($\tau_{max} = \alpha\tau_{max}^{We}(\dot{\varepsilon})$) Cost function =1665 | 4374 | +36% | 0.764 | -24% | 1.85 | +12.6% |

Table 4: Identifications performed on Experiment 3 (imposed constant displacement rate of $\dot{u} = 0.08\ mm/s$)

Concerning Identifications 2 and 3, the accommodation required on $\tau_{\text{max}}$ through $\alpha$ is much weaker in the case where the Weissenberg companion relation is used (+4.5% against -14% for eq. 12). Because in that case the strain rate variations are very important (up to $1.7 \, 10^{-2} \, s^{-1} \, in \, Fig. 1$) this may suggest that the Weissenberg approach (eq. 11) is more sounded that the mathematical fit (eq. 12). Still, these last results confirm the intrinsic character of the constitutive law proposed in Eq.6 for the HDPE semi-crystalline polymer, which is able to model tensile test in imposed displacement command mode, once a complementary information is joint to the model to account for strain-rate dependency.

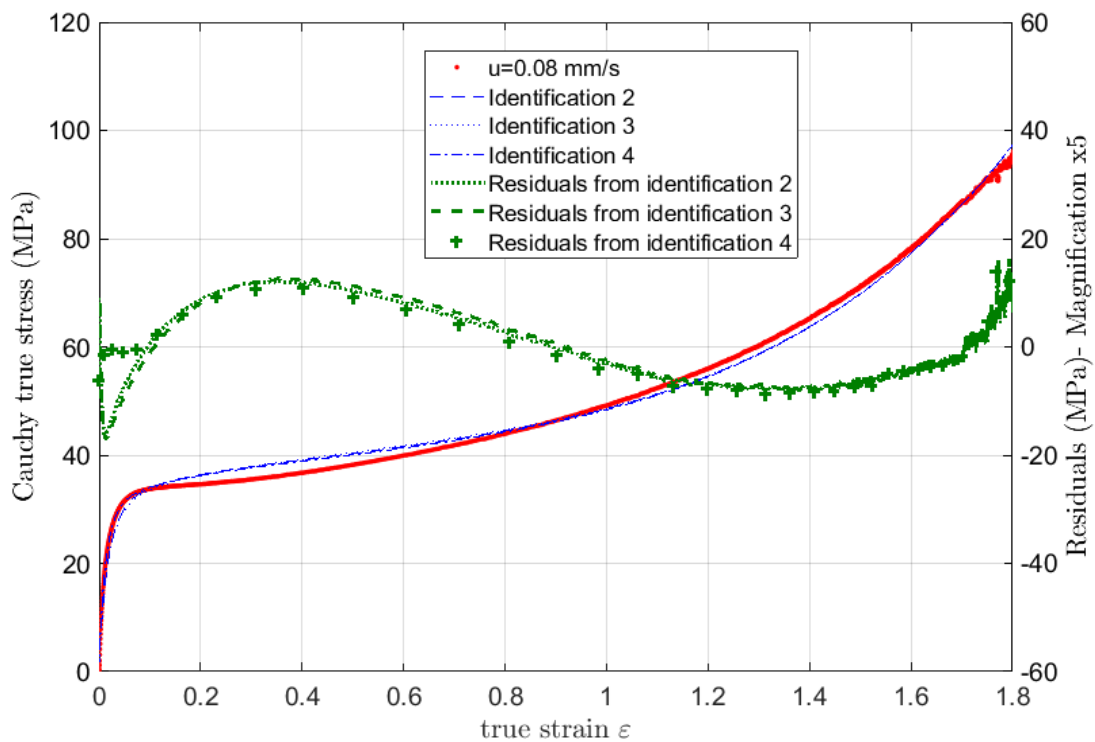

Figure 6: Comparison between tensile tests at constant displacement rate and direct model computation from eq.6 and companion strain-rate dependency for the relaxation spectrum (Eq.12).

### 6. Conclusion

The conclusion of this study is two folded. The prime conclusion was the objective of the paper : illustrating how the proof can be made that a proposed constitutive model is really intrinsic to the material. Following previous results reported in Blaise et al., 2016, it is shown that a proper application of theoretical concepts in P.E.P. allows the 'best as possible' estimation of the model parameters. These parameters if really intrinsic to the material should convey strain-rate dependency if tensile tests are performed at different strain-rates. This is an example but the method could be extended for example to account for temperature dependence. Once a 'companion' relation is derived for this strain-rate effect on the parameters, the probation test consists in considering tensile tests performed for varying strain-rates all along the tensile test - like for imposed displacement-rates - as a result of strain localization (or necking) appearing in SCP specimen. The pair of constitutive behavior's model and strain-rate dependency description relations is shown to be able to fairly well predict the material behaviors when the strain and strain-rates vary.

The secondary conclusion should possibly lead to better practices in mechanical characterization. The constant strain-rate tensile test, coupled to a properly designed theoretical behavior's law and a deep investigation of the parameter identifiability problem is very effective to produce meaningful material parameter estimations. It is a way which could allow evolution of international standards, knowing that the ones still recommended by the ASTM and ISO ones still remain for example, on the estimation of an instantaneous Young modulus as the slope of a linear approximation of the stress-strain curve in a normalized strain interval. In polymers, viscoelastic effects start as soon as a mechanical excitation is imposed on the specimen and a modulus estimation should take this into account.

**Acknowledgements** The authors gratefully acknowledge financial support from different research programs involving the CPER 'SusChemProc' of Grand-Est Region, the CNRS 'PEPS' programs 2018 and 2019, the CARNOT label for its specific financial support #ICEEL-2019.and the EMPP Research Axis of Lorraine University.

# References

André, S., Meshaka, Y., Cunat, C.: Rheological constitutive equation of solids: a link between models based on irreversible thermodynamics and on fractional order derivative equations. Rheologica Acta. **42**, 500-515 (2003)

André, S., Renault, N., Meshaka, Y., Cunat, C.: From the thermodynamics of constitutive laws to thermomechanical experimental characterization of materials: An assessment based on inversion of thermal images, Continuum Mechanics and Thermodynamics, **24**, 1-20 (2012).

Arruda, E.M., Boyce, M.C., A three-dimensional constitutive model for the large stretch behavior of rubber elastic materials, Journal of the Mechanics and Physics of Solids, **41**(2), 389 (1993)

Aster, R.C., Borchers, B., Thurber, C.H, Parameter Estimation and Inverse Problems, S$^{nd}$ ed., Academic Press. (2013)

Beck, J.V., Arnold, K.J., Parameter Estimation in Engineering and Science, John Wiley & Sons, New York (1977)

Blaise A., Andre S., Delobelle P., Meshaka Y., Cunat C., Advantages of a 3-parameter Reduced Constitutive Model for the Measurement of Polymers Elastic Modulus using Tensile Tests., Mechanics of Time-Dependent Materials, 20(4), 553-577 (2016).

Blaise, A., Baravian, C., André, S., Dillet, J., Michot, L.J., Mokso, R, .Investigation of the mesostructure of a mechanically deformed HDPE by synchrotron microtomography, Macromolecules, 43, 8143-8152 (2010).

Cunat, C., A thermodynamic theory of relaxation based on a distribution of non-linear processes. J. Non-Crystalline Solids. 131/133, 196-199 (1991)

Cunat, C., The DNLR approach and relaxation phenomena : part I : Historical account and DNLR formalism. Mech. Of Time-Depend. Mater.. 5, 39-65 (2001)

De Donder, T., Thermodynamic theory of affinity: A book of principle, Oxford University Press. (1936)

Faccio-Toussaint, E., Ayadi, Z., Pilvin, P., Cunat, C., Modeling of the Mechanical Behavior of a Nickel alloy by Using a Time-Dependent Thermodynamic Approach to Relaxations of Continuous Media.. Mech. Of Time-Depend. Mater. 5, 1-25 (2001)

Farge L., Boisse J., Dillet J., André S., Albouy P-A, Meneau F., WAXS study of the lamellar/fibrillar transition for a semi-crystalline polymer deformed in tension in relation with the evolution of volume strain, Journal of Polymer Science, B Polymer Physics, 53, 1470–1480 (2015).

Haward, R.N., Strain Hardening of Thermoplastics. Macromolecules. 26, 5860-5869 (1993)

Haward, R.N., Strain hardening of High Density Polyethylene. J. Polymer Sciences, Part B: Polymer Physics. 45, 1090-1099 (2007)

Lion, A., On the Thermodynamics of fractional damping elements. Continuum Mech. Thermodyn. 9, 83-96 (1997).

Maillet, D., André, S., Rémy, B., Degiovanni, A., Regularized parameter estimation through iterative rescaling (PETIR): an alternative to Levenberg-Marquardt's algorithm, (2013) HAL Id: hal-00867608 https://hal.archives-ouvertes.fr/hal-00867608.

Na, B.,. Lv, R Xu, W. Yu, P., Wang, K., Fu, Q., Inverse temperature dependence of strain hardening in ultrahigh molecular weight polyethylene: Role of lamellar coupling and entanglement density, J. Phys. Chem. B 111(46) 13206-13210 (2007).

Prigogine, I., Defay, R.: Thermodynamique chimique conformément aux méthodes de Gibbs et De Donder, Tomes I-II. Gauthier-Villars (1944-1946).

Sorvari, J., Hämäläinen J., Time integration in linear viscoelasticity – a comparative study. Mech. Time-Depend. Mater. 14, 307-328 (2010)

Tervoort, T. A.; Govaert, L. E., Strain-hardening behavior of polycarbonate in the glassy state J Rheol, 44, 1263, (2000).

Treloar, L.R.G., The Physics of Rubber Elasticity. Clarendon Press, Oxford, UK. (1975)

Van Melick, H.G.H., Govaert, L.E., Meijer, H.E.H., On the origin of strain hardening in glassy polymers, Polymer, 44(8) 2493-2502 (2003).

Walter, E, Pronzato, L, Identification of Parametric Models from Experimental Data Springer-Verlag  Heidelberg (1997).

Ye, J, Andre, S., Farge, L., Kinematic study of necking in a semi-crystalline polymer through 3D digital image correlation, Int. J. Solids & Structures, 59, 58-72 (2015).

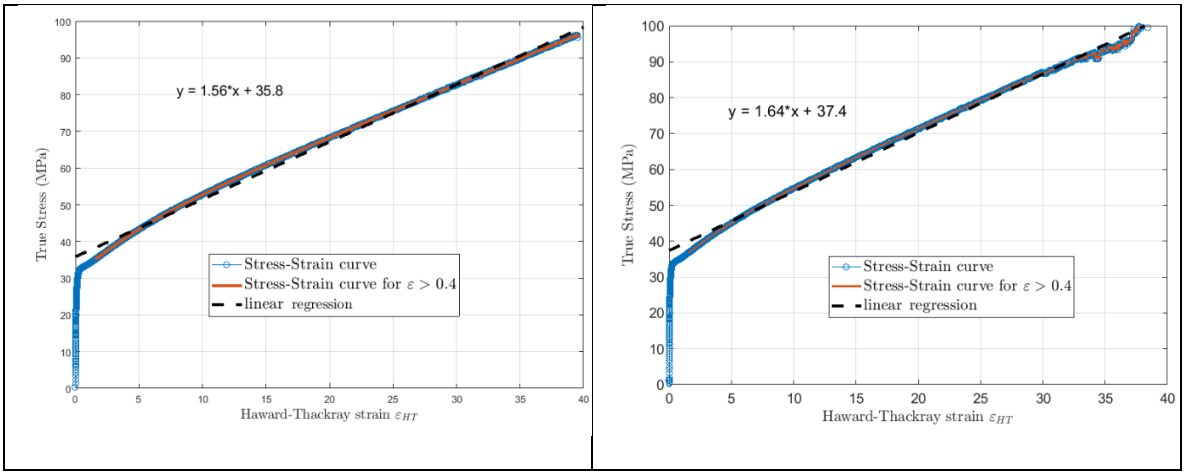

Figure S1: Stress-Strain curves represented for u=0.02mm/s (left) and u=0.08mm/s (right) as a function of the Haward-Thackray strain in abscissa. A linear regression on the datas for $\varepsilon > 0.4$ gives a direct estimation of the hardening modulus $G$.

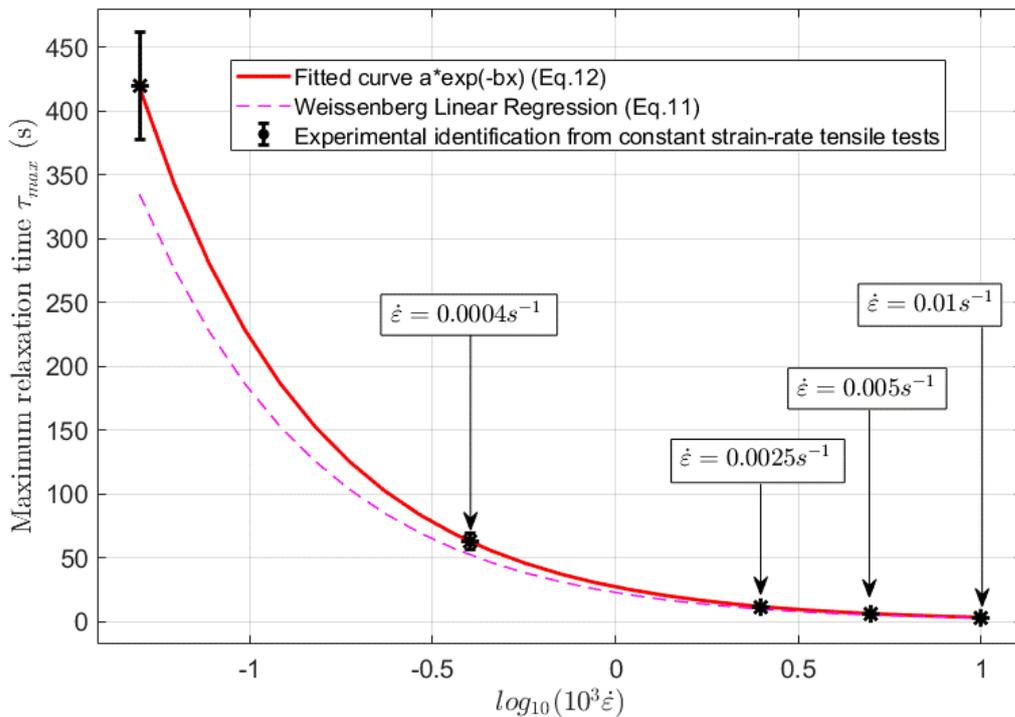

Figure S2: Maximum relaxation times identified experimentally from constant strain-rate tensile tests (Experiment1) plotted as function of strain rate, along with the two relations (Eq.11 and Eq.12 in dashed and solid lines) proposed for their description. Error bars correspond to a fictitious 10% uncertainty.

**Curve fitting process for the command/excitation signal**

When the excitation in the real experiment is applied on the grip displacement $(u, \dot{u})$, curve fitting of the excitation variables $(\varepsilon, \dot{\varepsilon})$ of the behavior's law is required and was performed under Matlab Curve fitting Toolbox 3.5.6. Note that very good fits are sought to in order to make the model computations precise and un-biased but there is no need of any explicative model. Mathematical forms can then be selected without any physical constraints. For this reason the curve fitting process is performed in two steps to enhance the quality.

A first fit is performed of the signal $\varepsilon(t)$. In many cases, the mathematical function used for the fit is of the form

$$\varepsilon^{fit}(t) = \frac{bt}{(t^c + d\, e^{at})} \quad \text{(A-1)}$$

An example is given in the figure S3a below where *texp* figures the experimental time and *xexp* the strain variable as measured with the DIC system.

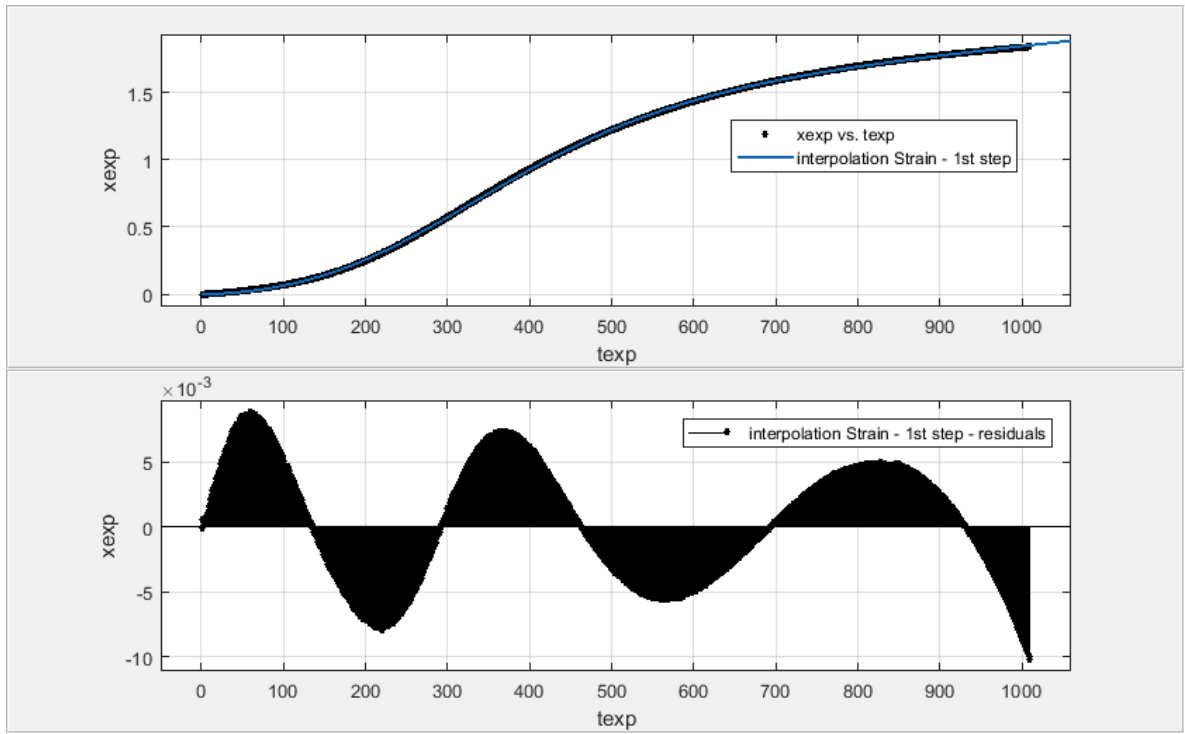

Figure S3: (a) upper plot of experimental strain versus time. (b) fitting residuals.

Residuals $r(t) = \varepsilon(t) - \varepsilon^{fit}(t)$ are calculated after this first step.

A second fit is performed on these residuals. Generally a series in sine function of a few terms is obviously appropriate.

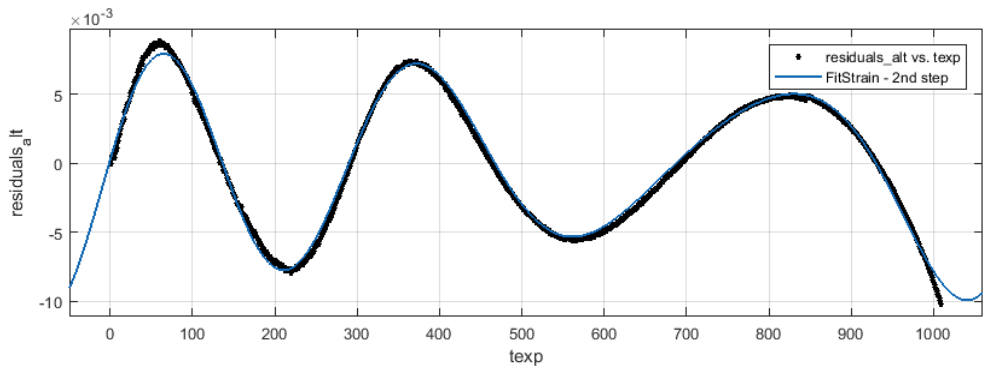

Figure S4 : Fit of the residuals under a series of sine functions.

Figure S5 plot the residuals after this second step. They are in this example of the order of 0.001 for a signal which varies from 0 to 1.8.

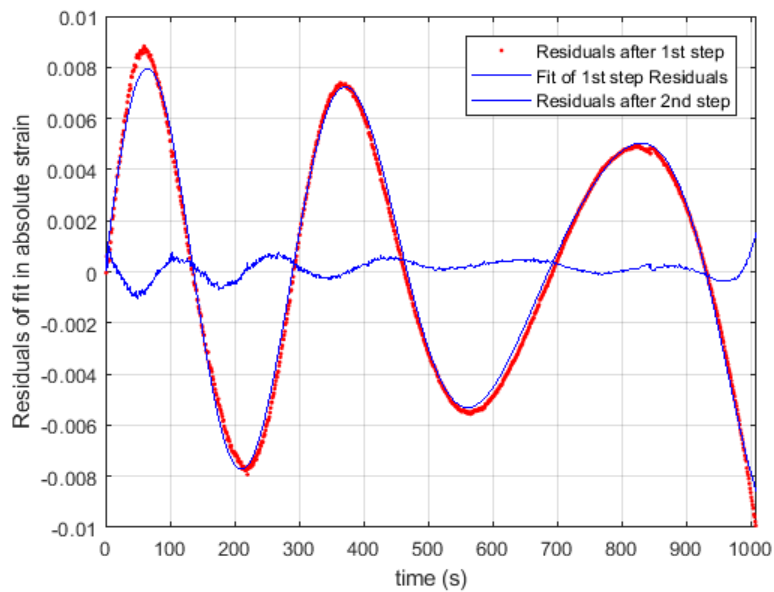

Figure S5 : fit of the 1st step residuals and residuals after this second step.

Then Figure S6 shows the strain fit that can be obtained for the real experiment at u=0.02mm/s

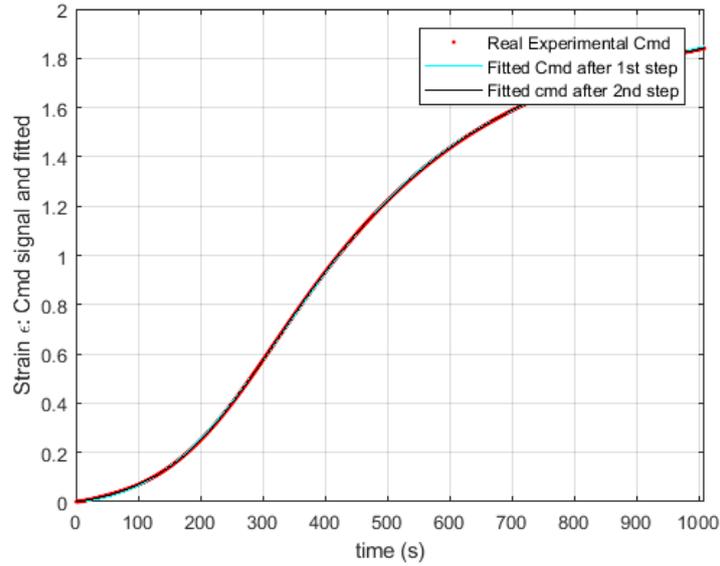

Figure S6 : Final curve fitting of the experimental strain signal.

The process can be repeated now on the experimental strain-rate signal $\dot{\varepsilon}$ which is obtained from a numerical differentiation. Figure S7 gives an example of the experimental strain-rate signal and its 1-step and final (2-step) fitted counterpart.

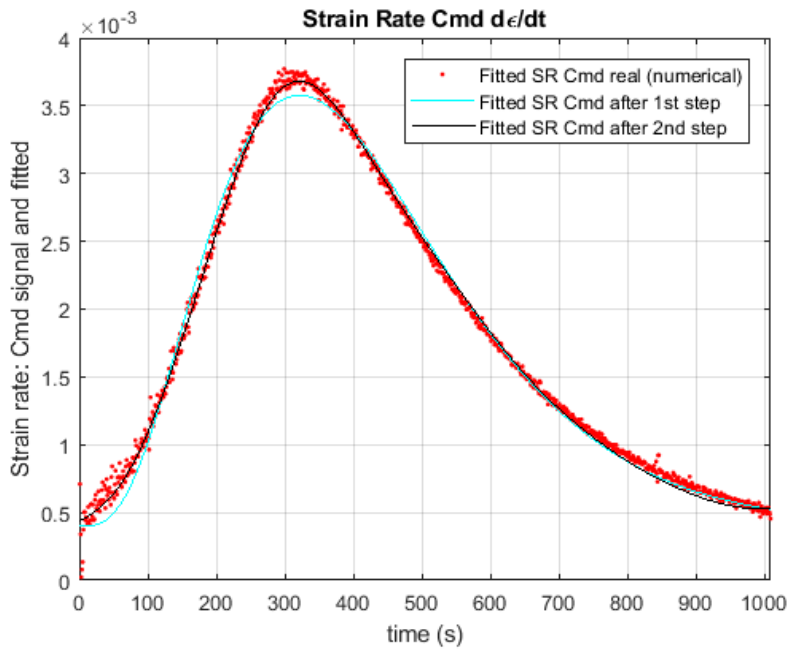

Figure S7 : Experimental and successively fitted strain-rate signal.

Note that having a good un-noisy signal for the strain-rate realized during the experiment, it is possible to compute the strain signal resulting from the displacement command by cumulative integration $\varepsilon(t) = \int_0^t \dot{\varepsilon}(u)\,\mathrm{d}u$ to check and validate the whole process. This is what is shown in figure S8 below.

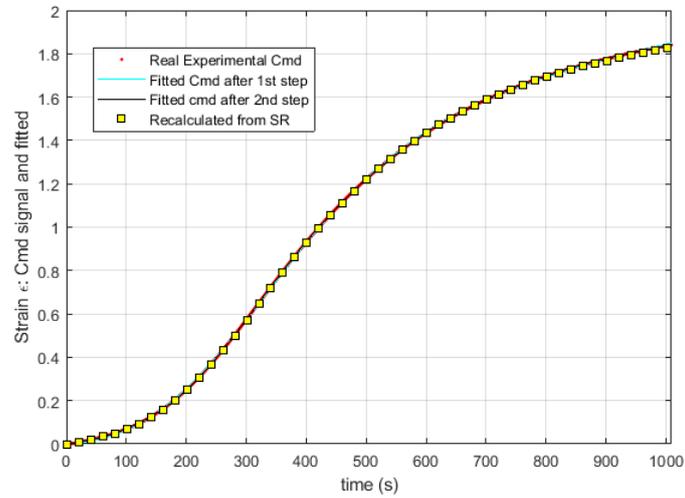

Figure S8 : Validation of the strain-rate fit with a numerical time integration to recover the strain signal.